\documentclass[ %
12pt,
a4paper,
ngerman,
english,
]{scrartcl} %

\usepackage{babel}              
\usepackage[utf8]{inputenc}   
\usepackage[T1]{fontenc}        
\usepackage[scaled]{helvet}     

\usepackage{amsmath}
\usepackage{amsfonts}
\usepackage{amssymb}
\usepackage{amsthm}    
\usepackage{graphicx}   
\usepackage{caption}
\usepackage{subcaption}
\usepackage[automark]{scrpage2} 
\usepackage{geometry} 
\usepackage[round,comma,sort]{natbib} 
\usepackage{color}
\usepackage{epstopdf}           
\usepackage{fancybox}           
\usepackage{tabu}
\usepackage{hhline}
\usepackage[draft]{hyperref}
\hypersetup{final}
\usepackage{cleveref}
\usepackage{makeidx}            
\usepackage{enumerate}          
\usepackage{doi}
\makeindex

\newif\ifPics %
\Picstrue 

\begin{document}

\title{Comparing seven variants of the Ensemble Kalman Filter: How
  many synthetic experiments are needed?} %
\author{Johannes Keller%
  \thanks{ %
    Corresponding Author: jkeller@eonerc.rwth-aachen.de, %
    Institute for Applied Geophysics and Geothermal Energy, %
    E.ON Energy Research Center, %
    RWTH Aachen University, %
    Aachen, %
    Germany %
  } %
  \and Harrie-Jan Hendricks Franssen%
  \thanks{ %
    Forschungszentrum J{\"u}lich GmbH, %
    Institute of Bio- and Geosciences, %
    IBG-3 (Agrosphere), %
    J{\"u}lich, %
    Germany %
  } %
  \thanks{ %
    Centre for High-Performance Scientific Computing in Terrestrial
    Systems (HPSC-TerrSys), %
    Geoverbund ABC/J, %
    J{\"u}lich, %
    Germany %
  } %
  \and Gabriele Marquart%
  \footnotemark[1] %
} %
\date{} %
\maketitle

Accepted for publication in Water Resources Research. Copyright 2018
American Geophysical Union. Further reproduction or electronic
distribution is not permitted.

\newpage

\begin{abstract}
  The Ensemble Kalman Filter (EnKF) is a popular estimation technique
  in the geosciences. %
  It is used as a numerical tool for state vector prognosis and
  parameter estimation. %
  The EnKF can, for example, help to evaluate the geothermal potential
  of an aquifer. %
  In such applications, the EnKF is often used with small or medium
  ensemble sizes. %
  It is therefore of interest to characterize the EnKF behavior for
  these ensemble sizes. %
  For seven ensemble sizes (50, 70, 100, 250, 500, 1000, 2000) and
  seven EnKF-variants (Damped, Iterative, Local, Hybrid, Dual, Normal
  Score and Classical EnKF), we computed 1000 synthetic parameter
  estimation experiments for two set-ups: a 2D tracer transport
  problem and a 2D flow problem with one injection well. %
  For each model, the only difference among synthetic experiments was
  the generated set of random permeability fields. %
  The 1000 synthetic experiments allow to calculate the pdf of the
  RMSE of the characterization of the permeability field. %
  Comparing mean RMSEs for different EnKF-variants, ensemble sizes and
  flow/transport set-ups suggests that multiple synthetic experiments
  are needed for a solid performance comparison. %
  In this work, 10 synthetic experiments were needed to correctly
  distinguish RMSE differences between EnKF-variants smaller than
  10\%. %
  For detecting RMSE differences smaller than 2\%, 100 synthetic
  experiments were needed for ensemble sizes 50, 70, 100 and 250. %
  The overall ranking of the EnKF-variants is strongly dependent on
  the physical model set-up and the ensemble size. %
\end{abstract}

\section{Introduction}
\label{sec:introduction}

Modeling groundwater and heat flow is a key step for many subsurface
applications like the evaluation and possible utilization of
geothermal systems. %
Subsurface flow and heat transport processes are strongly influenced
by the permeability of the porous rock. %
Therefore, the inverse estimation of permeabilities has been studied
extensively in the groundwater literature of the last decades
\citep[e.g.,][]{Kitanidis1983, Carrera1986, RamaRao1995,
  GomezHernandez1997, Hendricks2003, Chen2006, Nowak2009,
  Kurtz2014}. %
Since the 1980's a strong focus has been laid on the use of stochastic
methods for solving inverse problems. %
They allow to quantify the uncertainty associated with inversely
estimated parameters. %
Justification for the stochastic methods is provided by probability
theory, for instance using a Bayesian framework \citep{Chen2003}. %
However, nonlinear models of the subsurface with many unknown
parameters pose challenges to stochastic inverse modeling. %
Sequential methods, such as the Ensemble Kalman Filter
\citep[EnKF,][]{Evensen2003}, were found to be relatively efficient
for these nonlinear models, still allowing for robust uncertainty
quantification in spite of many unknowns. %
In the EnKF, uncertainty is captured by a large number of stochastic
realizations. %
The size of this ensemble is a crucial influencing factor for filter
performance. %
It was observed in several studies that small ensemble sizes lead to
problems such as spurious correlations, which can result in filter
inbreeding and ultimately filter divergence
\citep[e.g.,][]{Hamill2001}. %

In the following, typical ensemble sizes used in the Ensemble Kalman
Filter literature are reviewed. %
In some applications of the EnKF in the atmospheric sciences
\citep{Houtekamer1998, Anderson2001, Hamill2000, Kalnay2002} ensemble
sizes up to 500 are used. %
However, most of the time, model complexity restricts ensemble sizes
to less than 100. %
In the field of land surface modeling, where the EnKF is mostly used
for state vector estimation, ensemble sizes smaller than 100 are
common \citep[e.g.,][]{Reichle2002, DeLannoy2016}. %
If we consider parameter estimation studies in surface and subsurface
hydrology, in many cases ensemble sizes smaller than 250 are used
\citep[e.g.,][]{Lorentzen2003, Naevdal2005, Vrugt2005, Chen2006,
  Shi2014, Baatz2017}. %
More uncommon are ensemble sizes larger than 500
\citep[e.g.,][]{Devegowda2010, Vogt2012}.

Different variants of the EnKF have been proposed to reduce problems
linked to small ensemble sizes, such as large fluctuations in the
sampled model covariances which might result in filter inbreeding and
filter divergence. %
In this study, we compare the following methods: damping of the EnKF
\citep{Hendricks2008}, Local EnKF \citep{Hamill2001}, Hybrid EnKF
\citep{Hamill2000}, Dual EnKF \citep{Moradkhani2005, ElGharamti2013},
Iterative EnKF \citep{Sakov2012} and Normal Score EnKF
\citep{Zhou2011_3, Schoniger2012, Li2012}. %
Implementation details of these algorithms are given in Section
\ref{sec:methods}. %
Of course, many methods exist that are not compared in this study. %
Two examples are modern localization approaches \citep{Chen2010,
  Emerick2011} and methods using decompositions of probability
distributions into sums of Gaussian distributions \citep{Sun2009_2,
  Liu2016}. %

To assess the benefits of new EnKF-variants or to compare known
variants applied to different geoscience models, it is important to
have meaningful comparison methods. %
Ideally, we would like to compare the inverse solution with the exact,
correct inverse solution solving Bayes theorem. %
However, an analytical solution is not possible for realistic problems
and a numerical solution by Markov Chain Monte Carlo methods is
extremely expensive. %
Therefore, for large inverse problems the correct inverse solution
will be unknown. %
Three typical remedies, which can be found in the literature, include:
Comparing estimation results with a synthetic reference; comparing
results of a target variable with measurements (e.g. oil production in
reservoir engineering); or comparing small-ensemble runs with runs
using a larger ensemble. %
In all cases, EnKF-variants are usually compared by evaluating a small
number of synthetic experiments. %
A synthetic experiment is defined here as the comparison of multiple
data assimilation methods by applying them to the same parameter
estimation problem with fixed synthetic reference field, fixed
ensemble size and fixed measurements. %

There are several examples of EnKF comparison studies in the
hydrologic sciences. %
When performance measures such as the RMSE or the MAE were considered,
differences between methods often were smaller than 10\%
\citep[e.g.,][]{Moradkhani2005, Camporese2009, Sun2009_2,
  Hendricks2009_2, Zhou2011_3, ElGharamti2014_3, ElGharamti2014,
  ElGharamti2015, Liu2016}. %
The comparison of EnKF algorithms was often based on less than 10
synthetic experiments. %
Moreover, with the strict definition of synthetic experiment given
above (comparing data assimilation methods for identical sets of
e.g. ensemble size, initial parameter fields), most studies were based
on single synthetic experiments. %
Some studies evaluated multiple synthetic experiments of test cases. %
\citet{Moradkhani2005} employed 500 synthetic experiments of ensemble
size 50, \citet{Sun2009} used 16 synthetic experiments and
\citet{Schoniger2012} computed 200 test cases. %
In this paper, we consider two simple subsurface models and aim to
evaluate the number of synthetic experiments needed to distinguish
performances of EnKF-variants for each of these two models. %

The central motivation of this study is to characterize the random
component inherent in the comparison of the performance of different
EnKF-variants. %
Such comparisons are often based on small or medium size ensembles and
one or a few synthetic experiments. %
The random component results from the limited ensemble of initial
parameter fields, as well as from measurement errors and the
associated sampling fluctuations. %
Applying the EnKF-variants to large models, one is computationally
restricted to small ensemble sizes and few synthetic experiments,
which is why often final results are still subject to considerable
uncertainty and a function of specific filter settings. %
For small models, such as the ones presented in this paper, it may be
possible to suppress the most notorious problems of undersampling by
sufficiently enlarging the ensemble size. %
Nevertheless, knowing how EnKF-methods perform on small models might
help in deciding which EnKF-method to apply to a large model with
similar features. %
However, it cannot be guaranteed that the same EnKF-methods which work
well on small models would also work well on larger models. %
Increasing computer power provides the opportunity to gain this
knowledge by computing many synthetic experiments. %
In this study, we monitor EnKF performance by running large numbers of
synthetic experiments for small and medium ensemble sizes. %
We hypothesize that randomness is non-negligible in the evaluation of
a small number of synthetic experiments. %
The influence of the random component is characterized by calculating
root mean square errors between estimated permeability fields and the
synthetic true field for each synthetic experiment. %
For 1, 10 and 100 synthetic experiments, it is evaluated how strongly
the RMSEs deviate from the mean over all 1000 synthetic experiments. %
Finally, we determine the performance difference between EnKF-variants
(in terms of RMSE) which can be detected with a given number of
synthetic experiments (1, 10 or 100). %

In Section \ref{sec:methods}, the model equations and the EnKF-methods
are introduced. %
Section \ref{sec:design-syn-exp} presents the design of the synthetic
experiments. %
Subsequently, the tools for comparison of EnKF-methods are
explained. %
The results of the different sets of synthetic experiments for the two
different flow/transport configurations, the different EnKF-variants
and ensemble sizes, are presented in Section
\ref{sec:discussion-results}. %
The results are also discussed in that section. %
Finally, in Section \ref{sec:conclusion}, our conclusions are
presented. %

\section{Methods}
\label{sec:methods}

\subsection{Governing equations and their solution}
\label{sec:gov-eqns-sol}

The transient groundwater flow equation considered in this study is
given by: %
\begin{equation}
  \label{groundwater-flow}
  S_{s}\frac{\partial h}{\partial t} = \nabla \cdot
  \mathbf{v},
\end{equation}
where $h$ $[L]$ denotes the hydraulic head and $t$ $[T]$ denotes
time. %
$S_{s}$ $[L^{-1}]$ is the specific storage.  Simulation of groundwater
flow is based on Darcy's law for the computation of the groundwater
flow velocity $\mathbf{v}$ $[LT^{-1}]$: %
\begin{equation}
  \label{darcy-velocity}
  \mathbf{v} = \frac{\rho_{f}g}{\mu_{f}}\mathbf{k}\cdot \vec{\nabla}h.
\end{equation}
Here, $g$ $[LT^{-2}]$ is the gravitational constant, $\rho_{f}$
$[ML^{-3}]$ is water density, $\mu_{f}$ $[ML^{-1}T^{-1}]$ is the
dynamic viscosity of water and $\mathbf{k}$ $[L^{2}]$ is the hydraulic
permeability tensor (\citet{Bear1975}). %
In this study, the water density $\rho_{f}$, the gravitational
constant $g$ and the dynamic viscosity of water $\mu_{f}$ are assumed
constant in space and time. %
The hydraulic permeability tensor $\mathbf{k}$ is the aquifer
parameter of interest. %
In this study it is assumed to be isotropic. %
In the isotropic case, a scalar permeability $K$ specifies the full
tensor. %

The equation for solute transport is given by: %
\begin{equation}
  \label{eq:tracer-transport}
  \varphi \frac{\partial c}{\partial t} =
  \nabla
  \left(
    \mathbf{D} \cdot \nabla c
  \right) -
  \mathbf{v} \cdot \nabla c,
\end{equation}
where $c$ $[ML^{-3}]$ is the concentration of the tracer. %
The porosity of the rock matrix $\varphi$ $[-]$ is assumed constant. %
In our model, the hydrodynamic dispersion tensor $\mathbf{D}$
$[L^{2}T^{-1}]$ has a smaller influence on concentration evolution
than the Darcy velocity. %

The numerical software SHEMAT-Suite can solve coupled transient
equations for groundwater flow, heat transport and reactive solute
transport \citep{Rath2006,Clauser2012}. %
In this study, we use it to simulate a conservative tracer experiment
for transient groundwater flow and another transient groundwater flow
case including an injection well. %
The temperature is constant during both simulations. %

The software SHEMAT-Suite \citep{Rath2006} uses the finite difference
method to solve the governing equations. %
They are solved implicitly \citep{Lynch2005, Huyakorn2012}. %

\subsection{EnKF}
\label{sec:enkf}

The Ensemble Kalman Filter (EnKF) is an ensemble-based data
assimilation algorithm derived from the classical Kalman Filter. %
The classical Kalman Filter sequentially updates a state vector and
its covariance matrix by optimally weighting model predictions and
measurements. %
The Kalman Filter provides optimal solutions for linear systems and
Gaussian statistics \citep{Kalman1960}. %

The Ensemble Kalman Filter is a Monte Carlo variant of the Kalman
Filter \citep{Evensen1994}. %
EnKF performs better than the classical Kalman Filter for non-linear
dynamics and can be applied to larger systems with many unknowns. %
Instead of calculating the model covariance matrix analytically, it is
approximated from an ensemble of model simulations. %
This ensemble should capture the main uncertainty sources relevant for
the model prediction. %
For groundwater flow and solute transport, the most important
uncertainty source is typically the rock permeability governing the
hydraulic conductivity. %
By augmenting the state vector by rock permeability, the EnKF can be
used for parameter estimation. %

The EnKF consists of three main steps: The forward simulation
(indicated by superscript $f$), the measurement equation and the
update equation (indicated by superscript $a$). %
During forward simulation, the model is applied to each realization. %
\begin{equation}
  \label{eq:kalman-forward}
  \mathbf{x}^{f}_{i,t_{j+1}} = M
  \left(
    \mathbf{x}_{i,t_{j}}^{a}
  \right) \qquad i \in
  \left\{
    1, \cdots, n_{e}
  \right\}, j \in
  \left\{
    1, \cdots, n_{t}
  \right\}.
\end{equation}
The vector $\mathbf{x}^{a}_{i,t_{j}} \in R^{n_{s}}$ holds the $i^{th}$
realization of states and parameters \emph{after} assimilation time
$t_{j}$. %
In the special case $j=1$, it holds the $i^{th}$ realization of the
\emph{initial} states and parameters. %
The vector's dimension $n_{s}$ is the sum of the number of states and
the number of parameters, $n_{e}$ is the number of ensemble members
and $n_{t}$ is the number of assimilation times. %
In our case, the model $M$ represents the solution of the groundwater
flow and solute transport equations for the hydraulic head and solute
concentration in $\mathbf{x}^{a}_{i,t_{j}}$. %
The parameters in $\mathbf{x}^{a}_{i,t_{j}}$ are left constant by
$M$. %
The vector $\mathbf{x}^{f}_{i,t_{j+1}} \in R^{n_{s}}$ holds the
$i^{th}$ realization of states and parameters \emph{before}
assimilation at time $t_{j+1}$. %
For simplicity, we will drop the time index for the rest of this
section. %

The ensemble of measurements is given by: %
\begin{equation}
  \mathbf{d}_{i} = \mathbf{H}\mathbf{x}^{\mathrm{meas}} + \epsilon_{i} \qquad i \in
  \left\{
    1, \cdots, n_{e}
  \right\}.
\end{equation}
Observed measurement values
$\mathbf{y} = \mathbf{H}\mathbf{x}^{\mathrm{meas}} \in R^{n_{m}}$
(tracer concentrations or hydraulic heads from the reference model in
our specific set-ups) are perturbed by Gaussian noise
$\epsilon_{i} \in R^{n_{m}}$ with covariance matrix
$\mathbf{R} \in R^{n_{m} \times n_{m}}$ (\cite{Burgers1998}). %
Here, $n_{m}$ is the number of measurements at the time step under
consideration. %
In general, measurement values are connected to the states and
parameters of the model through the measurement operator
$\mathbf{H} \in R^{n_{m}\times n_{s}}$. %
In the end, we obtain a vector of perturbed measurements
$\mathbf{d}_{i} \in R^{n_{m}}$ for each realization. %

The EnKF update equation is given by: %
\begin{equation}
  \label{eq:kalman-update}
  \mathbf{x}_{i}^{a} = \mathbf{x}_{i}^{f} +
  \mathbf{K}
  \left(
    \mathbf{d}_{i}-\mathbf{H}\mathbf{x}_{i}^{f}
  \right), \qquad i \in
  \left\{
    1, \cdots, n_{e}
  \right\}.
\end{equation}
For each realization, the prediction from the forward simulation
$\mathbf{x}_{i}^{f}$ is compared with the perturbed measurement vector
$\mathbf{d}_{i}$. %
Then, $\mathbf{x}_{i}^{f}$ is updated according to the Kalman Gain
matrix $\mathbf{K} \in R^{n_{s} \times n_{m}}$, which is given by: %
\begin{equation}
  \label{eq:Kalman-gain}
  \mathbf{K} = \mathbf{P}_{e}\mathbf{H}^{T}
  \left(
    \mathbf{H}\mathbf{P}_{e}\mathbf{H}^{T} + \mathbf{R}
  \right)^{-1}.
\end{equation}
Here, $\mathbf{P}_{e} \in R^{n_{s} \times n_{s}}$ denotes the ensemble
covariance matrix of states and parameters. %
$\mathbf{K}$ favors updates if ensemble covariances $\mathbf{P}_{e}$
are large compared to the measurement uncertainty $\mathbf{R}$. %

Details on how to implement the EnKF can be found in
\citet{Evensen2003}. %
We extended this implementation for joint state-parameter updating
according to \citet{Hendricks2008}. %
The following subsections contain introductions to the variants of the
EnKF algorithm compared in this case study. %

\subsubsection{Damping}

A damping factor $0<\alpha\leq 1$ can be included in the EnKF to
counteract filter divergence \citep{Hendricks2008}. %
The damped assimilation step is given by: %
\begin{equation}
  \label{eq:damping}
  \mathbf{x}_{i}^{a} = \mathbf{x}_{i} + \alpha \mathbf{K}
  \left(
    \mathbf{d}_{i} - \mathbf{H}\mathbf{x}_{i}
  \right).
\end{equation}
In this study, $\alpha$ only dampens the parameter updates - the state
updates are kept undamped. %
For groundwater flow simulation, damping of the parameter update
reduces the impact of ensemble-based linearization between the
hydraulic head and hydraulic conductivity. %
The ensemble covariance matrix necessarily treats any relation between
two states or a state and a parameter as linear, but for flow in
heterogeneous media this relation is non-linear. %
Updating the parameters by smaller steps, more slowly approximating
the posterior values, is therefore expected to be more stable. %
In synthetic experiments it is commonly found that this also reduces
filter inbreeding and filter divergence \citep{Hendricks2008,
  Wu2011}. %

\subsubsection{Localization}

For small ensemble sizes, undersampling can lead to large fluctuations
of ensemble covariances. %
Even for locations which are far apart in space, non-zero covariances
may appear, the so-called spurious correlations
\citep{Houtekamer1998}. %
Localization methods reduce the effect of these spurious long-range
correlations on the filter update. %
To this end, a correlation matrix $\rho \in R^{n_{s}\times n_{m}}$
\citep{Gaspari1999} is multiplied elementwise with the first part of
the Kalman gain \citep{Hamill2001}: %
\begin{equation}
  \label{eq:localization-kalman-gain}
  \mathbf{K}_{loc} = \left[
    \rho \circ
    \left(
      \mathbf{P}_{e}\mathbf{H}^{T}
    \right)
  \right]
  \left(
    \mathbf{H}\mathbf{P}_{e}\mathbf{H}^{T}+\mathbf{R}
  \right)^{-1}.
\end{equation}
Typically, a characteristic length scale $\lambda$ is associated with
the correlation matrix. %
For this study, both $\rho$ and its length scale are taken from
\citet{Gaspari1999}. %
An entry of the correlation matrix $\rho$ is a function of the
distance $d$ between two locations and the correlation length scale
$\lambda$. %
It is given by: %
\begin{equation}
  \rho(d,a =\sqrt{\frac{10}{3}}\lambda) =
  \left\{
    \begin{array}{ll}
      -
      \frac{ \left(\frac{d}{a}\right)^{5} }{ 4 }
      +
      \frac{ \left(\frac{d}{a}\right)^{4} }{ 2 }
      +
      \frac{ 5\left(\frac{d}{a}\right)^{3} }{ 8 }
      +
      \frac{ 5\left(\frac{d}{a}\right)^{2} }{ 3 }
      +
      1,
      & 0 \leq \frac{d}{a} < 1 \\[8pt]
      \frac{ \left(\frac{d}{a}\right)^{5} }{ 12 }
      +
      \frac{ \left(\frac{d}{a}\right)^{4} }{ 2 }
      +
      \frac{ 5\left(\frac{d}{a}\right)^{3} }{ 8 }
      +
      \frac{ 5\left(\frac{d}{a}\right)^{2} }{ 3 }
      -
      5\frac{d}{a}
      +
      4
      -
      \frac{ 2 }{ 3\frac{d}{a} },
      & 1 \leq \frac{d}{a} < 2 \\[8pt]
      0,
      & 2 \leq \frac{d}{a}.
    \end{array}
  \right.
\end{equation}
The parameter $a$ is a multiple of $\lambda$ adapted to the functional
form of $\rho$. %
The objective of $\rho$ is to approximate a two-dimensional Gaussian
bell curve: %
\begin{equation}
  G(d,\lambda) = \exp
  \left(
    -\frac{d^{2}}{2\lambda^{2}}
  \right).
\end{equation}
Contrary to the case of Gaussian correlations, $\rho$ is zero for
distances greater than $2a$. %

\subsubsection{Hybrid EnKF}

For Hybrid EnKF \citep{Hamill2000}, the covariance matrix is chosen as
a sum of the usual ensemble covariance matrix and a static background
covariance matrix: %
\begin{equation}
  \label{eq:hybrid-covariance}
  \mathbf{P}_{\mathrm{hybrid}} = \beta \mathbf{P}_{e} + (1-\beta) \mathbf{P}_{\mathrm{static}}.
\end{equation}
The factor $0\leq \beta \leq 1$ determines the weight assigned to the
ensemble covariance matrix. %
The static background covariance matrix represents prior knowledge
about the geology and physics of the model. %

\subsubsection{Dual EnKF}

Another method entering the comparison is Dual EnKF by
\citet{Moradkhani2005} and \citet{Wan2001}. %
The state vector $\mathbf{x}^{f}_{i}$ is split into two parts:
$\mathbf{x}^{f}_{s,i}$ contains the state variables and
$\mathbf{x}^{f}_{p,i}$ contains the parameters. %
When the filter reaches a measurement time, \emph{only the parameters}
are updated according to %
\begin{equation}
  \label{eq:dual-param-update}
  \mathbf{x}^{a}_{s,i} = \mathbf{x}^{f}_{s,i} \qquad
  \mathbf{x}^{a}_{p,i} = \mathbf{x}^{f}_{p,i} + \mathbf{K}_{p}
  \left(
    \mathbf{d}-\mathbf{H}_{p}\mathbf{x}^{f}_{p,i}
  \right).
\end{equation}
$\mathbf{K}_{p}$ and $\mathbf{H}_{p}$ are the parts of the Kalman gain
and measurement matrix projected onto the parameter space. %
After the parameter update, the forward simulation is recalculated
using the updated parameters. %
In the second and final updating step, \emph{only the states} are
updated according to %
\begin{equation}
  \label{eq:dual-state-update}
  \mathbf{x}^{a,2}_{s,i} = \mathbf{x}^{f,2}_{s,i} + \mathbf{K}_{s}
  \left(
    \mathbf{d}-\mathbf{H}_{s}\mathbf{x}^{f,2}_{s,i}
  \right)\qquad
  \mathbf{x}^{a,2}_{p,i} = \mathbf{x}^{f,2}_{p,i}= \mathbf{x}^{a}_{p,i}.
\end{equation}
The matrices $\mathbf{K}_{s}$ and $\mathbf{H}_{s}$ are projected onto
the space of state variables. %
Note that the same measurement values are used for both the parameter
and the state variable update. %
The whole procedure is repeated for each assimilation time. %

\subsubsection{Normal Score EnKF}
\label{sec:normal-score-enkf}

Normal Score EnKF \citep[NS-EnKF,][]{Zhou2011_3} was developed to
handle non-Gaussian probability distributions inside an EnKF
framework. %
The method inherits its name from the Normal Score transform
\citep{Goovaerts1997,Journel1978, Deutsch1992}. %
The ensemble of states, parameters and measurement values are
transformed before assimilation starts. %
The transform uses the cumulative distribution functions to turn the
ensemble into one of a normalized Gaussian pdf. %
\begin{equation}
  \label{eq:normal-score-transform}
  NS(x^{f}_{i}) := G^{-1}(F(x^{f}_{i})).
\end{equation}
Here, $x^{f}_{i}$ is the $i^{th}$ realization of a single component of
the state and parameter vector $\mathbf{x}^{f}_{i}$. %
$F$ is the cumulative distribution function of $x_{i}$ and $G$ is the
cumulative distribution function of the Gaussian pdf with zero mean
und unit standard deviation. %
The transform largely preserves the correlation structure of the
variables by keeping intact the relative order of the ensemble
members. %
The EnKF update is carried out in terms of the transformed values. %
After the update, states and parameters are back-transformed %
\begin{equation}
  \label{eq:normal-score-backtransform}
  NS^{-1}(x^{a}_{i}) := F^{-1}(G(x^{a}_{i})).
\end{equation}
While the back-transform is similar to the transform, a complication
arises, since $F$ is only an ensemble approximation (step-function) of
a cumulative distribution function. %
If $G(x^{a}_{i})$ falls into the range of the ensemble, $F^{-1}$ is
interpolated between the values of the two closest ensemble members. %
When updated ensemble members are located outside the range of the
ensemble, it is less trivial to perform the back-transform. %
Since $G(x^{f}_{i}) = \frac{i-0.5}{n_{e}}$, this corresponds to
either %
\begin{equation}
  \label{eq:normal-score-extrapolation}
  G(x^{a}_{i})<\frac{0.5}{n_{e}} \qquad \mathrm{or} \qquad
  G(x^{a}_{i})>\frac{n_{e}-0.5}{n_{e}}.
\end{equation}
In these cases, the following extrapolation method is used to
calculate $F^{-1}$ for outliers. %
Two artificial support points $x^{s}_{0}$ and $x^{s}_{1}$ are selected
and define the extreme values of the cumulative function: %
\begin{equation}
  \label{eq:normal-score-support-points}
  G(x^{s}_{0})=0 \qquad \qquad
  G(x^{s}_{1})=1.
\end{equation}
The distance between an artificial support point and the neighboring
support point of the cumulative function is set to three times the
spread of the original support points of $F^{-1}$. %
If one of the $x^{a}_{i}$ still lies outside of the support points, it
is moved to the closest support point. %
Once the distributions are properly back-transformed, the computation
of the forward model continues. %

\subsubsection{Iterative EnKF}

The iterative version of the EnKF (IEnKF) is inspired by the
exposition in \citet{Sakov2012}. %
Iteration is introduced to the EnKF to mitigate problems related to
the non-linearity of model equations. %
In our version, after an update of parameters by EnKF, the simulation
restarts from the beginning using the newly updated parameters as
input (in contrast, Dual EnKF always restarts from the previous
update). %
One drawback of this method is its computational demand. %
Let $T$ be the computing time of a normal EnKF run. %
Then, we can derive estimates for the computing times of Dual and
Iterative EnKF. %
For Dual EnKF, every update and forward computation is carried out
twice, thus, we expect a computing time of $T_{\mathrm{Dual}}=2T$. %
Assuming equidistant assimilation times for Iterative EnKF, the
computing time depends on the number of assimilation times $n_{T}$:
$T_{\mathrm{Iterative}}=T(\frac{1}{n_{T}}+\frac{2}{n_{T}}+\cdots+1)
=T\cdot\frac{n_{T}+1}{2}$. %
For the synthetic study in this paper ($n_{T}=100$), the ratio between
computation times amounts to
$\frac{T_{\mathrm{Iterative}}}{T_{\mathrm{Dual}}}=\frac{n_{T}+1}{4} \,
\tilde{=} \, 25$. %

\section{Design of the synthetic experiments}
\label{sec:design-syn-exp}

This section deals with the two set-ups for synthetic experiments
computed in this study, a forward model combining groundwater flow and
solute transport and a groundwater flow model with an injection
well. %

\subsection{Tracer model}
\label{sec:forward-model}

The 2D subsurface model for groundwater flow and solute transport is
based on a grid consisting of $31\,\times\,31$ cells of size
$2\, \mathrm{m} \times 2\, \mathrm{m}$. %
These $961$ cells define a square simulation domain of size
$62\, \mathrm{m} \times 62\, \mathrm{m}$. %
The governing equations are solved for a simulation time of $1200$
days divided into $200$ equal intervals. %

Boundary and initial conditions include a prescribed hydraulic head of
$11\, \mathrm{m}$ at the southern boundary and $10\, \mathrm{m}$ at
the northern boundary, creating a flow from south to north. %
The two remaining boundaries are impermeable. %
A tracer concentration of $80\times 10^{-3}\, \mathrm{mol}/\mathrm{l}$
is prescribed on the southern boundary; at the northern boundary we
set the concentration to
$60\times 10^{-3} \, \mathrm{mol}/\mathrm{l}$. %
The last value also serves as initial concentration throughout the
model domain leading to solute transport from south to north. %

The tracer is subjected to slow diffusion with a diffusion coefficient
of $1.5\times10^{-9}\, \mathrm{m}^{2}/\mathrm{s}$. %
The small diffusion coefficient ensures that the temporal evolution of
the tracer concentration is almost completely determined by
advection. %
The fluid is water with its standard properties. %
The porosity of the rock is $10\%$. %
At two locations with coordinates
$\left(19\, \mathrm{m},31\, \mathrm{m}\right)$ and
$\left(43\, \mathrm{m},31\, \mathrm{m}\right)$, the tracer
concentration is recorded every $12$ days, summing up to $100$
measurement times in total. %

\subsection{Well model}
\label{sec:ground-flow}

The 2D groundwater flow well model is based on a grid consisting of
$31\,\times\,31$ cells of size
$20\, \mathrm{m} \times 20\, \mathrm{m}$ resulting in a model domain
of $620\, \mathrm{m} \times 620\, \mathrm{m}$. %
The governing equations are solved for a simulation time of $18$ days
divided into $1200$ equal intervals. %
Boundary and initial conditions include a prescribed hydraulic head of
$11\, \mathrm{m}$ at a central well at coordinates
$\left(310\, \mathrm{m},310\, \mathrm{m}\right)$. %
All remaining boundary conditions and initial conditions are set to a
hydraulic head of $10\, \mathrm{m}$. %
Again, the fluid is water with its standard properties and the
porosity of the rock is $10\%$. %
In this model, there are $48$ measurement locations located along a
regular $7\times7$ grid excluding the central well position. %
Hydraulic heads are recorded at $60$ measurement times every $7$ hours
and $12$ minutes. %
The numerical settings for the solution of the linear system of
equations are identical for both the groundwater flow and mass
transport equation in the two cases. %
In the Picard iteration, we demand a relative tolerance of
$1\times 10^{-10}$. %
The linear solver BiCGStab is called with two termination criteria: An
accepted relative error of $1\times 10^{-14}$ and a maximum number of
$500$ iterations. %

\subsection{Simulation of permeability fields}
\label{sec:stoch-simul}

\begin{figure}
  \centering
  \includegraphics[trim = 50 350 50 300, clip,
    width=\columnwidth]{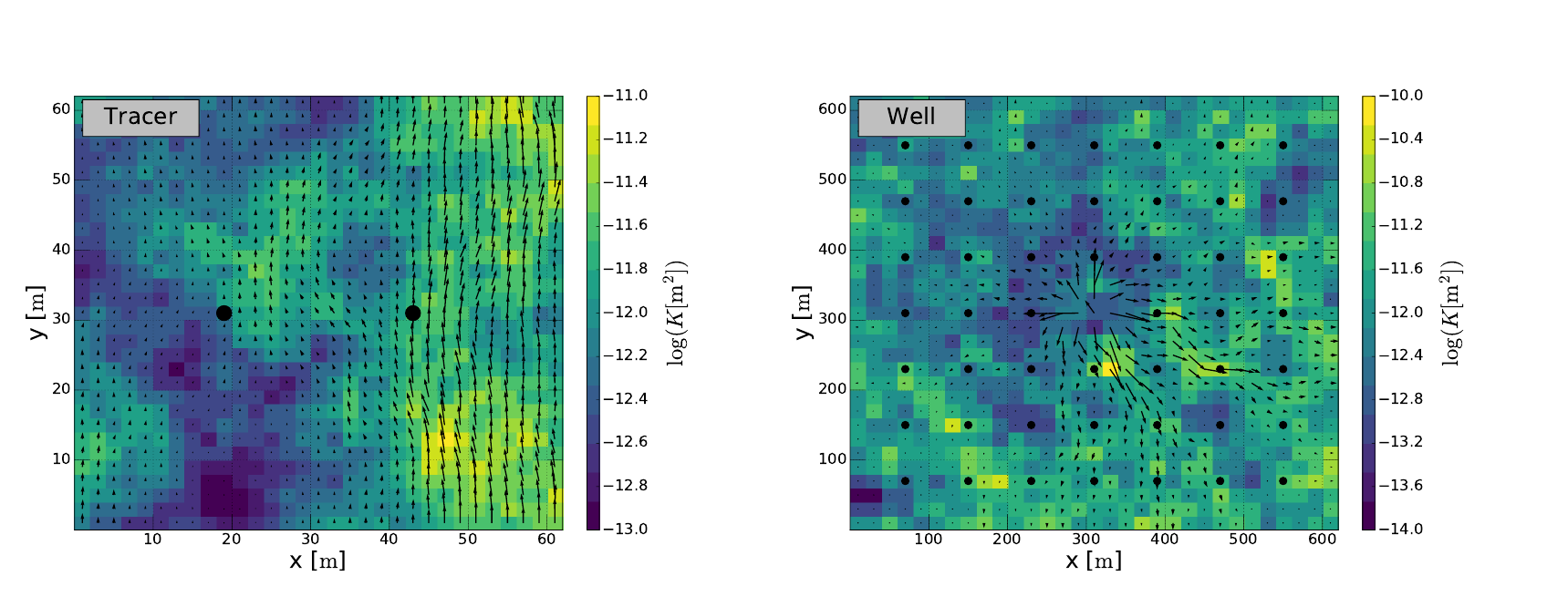}
    \caption{The logarithmic reference permeability fields for the
      tracer model and the well model with groundwater flow vectors.
      Measurement locations are depicted as black circles.}
  \label{fig:true-perm}
\end{figure}%

The heterogeneous synthetic reference permeability fields, displayed
in Figure \ref{fig:true-perm}, are generated by sequential
MultiGaussian simulation using the software SGSIM
\citep{Deutsch1992}. %
The following spherical correlation function is used: %
\begin{equation}
  \label{eq:spherical-correlation}
  \rho(d) = 1-
  \left(
    \frac{d}{a}
  \right)\cdot
  \left[
    \frac{3}{2} - \frac{1}{2}\cdot
    \left(
      \frac{d}{a}
    \right)
  \right].
\end{equation}
Here, $d$ denotes the Euclidean distance between two grid cells and
$a$ is the correlation length of the permeability field. %
No nugget effect is considered for the generation of the permeability
field. %
Both permeability fields are generated using mean
$-12.0\, \log_{10}(K[\mathrm{m}^{2}])$ and standard deviation
$0.5\, \log_{10}(K[\mathrm{m}^{2}])$. %
Correlation lengths of $a = 50\, \mathrm{m}$ for the tracer model and
$a = 60\, \mathrm{m}$ for the well model are used. %
Note that, relative to the full grid size, the correlation length of
the well model is smaller by one order of magnitude. %

The ensembles of initial permeability fields for the $1000$ synthetic
experiments (for each physical problem set-up, each ensemble size and
each EnKF-variant), which we use to compare performance, are generated
by SGSim using one set of input parameters, but varying random
seeds. %
A mean logarithmic permeability of
$-12.5\, \log_{10}(K[\mathrm{m}^{2}])$ (compared to the reference
fields with $-12.0\, \log_{10}(K[\mathrm{m}^{2}])$) and a standard
deviation of $0.5\, \log_{10}(K[\mathrm{m}^{2}])$ (identical to the
reference fields) are used. %
Consequently, the EnKF algorithms have to adjust the overall mean of
the permeability fields as well as their spatial variability. %

\subsection{EnKF setup}
\label{sec:enkf-input}

The EnKF state vector includes permeability, hydraulic head and, for
the tracer model, concentration. %
Permeability is the target parameter of the estimation. %
Concentration or hydraulic head values serve as the observations which
drive the estimation. %
Measurement noises are assumed constant:
$\sigma_{\mathtt{c}}= 7.1\times 10^{-3}\mathrm{mol}/\mathrm{l}$ for
concentration measurements and
$\sigma_{\mathtt{h}}= 5\times 10^{-2}\mathrm{m}$ for hydraulic head
measurements. %
These noises resemble the uncertainty of typical measurement
devices. %

Most EnKF-methods require the specification of additional
parameters. %
The damping factor of Damped EnKF is set to $\alpha=0.1$ (compare
\citep{Hendricks2008}). %
In Local EnKF, the length scale $\lambda$ is set to $150\, \mathrm{m}$
for the well model and the tracer model. %
For Hybrid EnKF, a constant diagonal background covariance matrix is
used. %
The value of $\beta$ used in this study is $0.5$. %
Results of synthetic experiments, for which $\lambda$ and $\beta$ were
varied, can be found in Section \ref{sec:vari-enkf-method-params}. %

In this study a series of EnKF data assimilation experiments, called
synthetic experiments, is performed. %
For each of the methods presented in Section \ref{sec:enkf}, seven
ensemble sizes (50, 70, 100, 250, 500, 1000 and 2000 realizations) are
tested. %
For the four smaller ensemble sizes, 1000 synthetic experiments are
carried out. %
For the larger ensemble sizes, due to computational limitations, only
100 synthetic experiments are computed. %

\subsection{Performance evaluation}
\label{sec:performance-eval}

For a single synthetic experiment with given EnKF-method and ensemble
size, the root mean square error is given by %
\begin{equation}
  \mathrm{RMSE}  = \sqrt{
    \frac{1}{n_{g}}\sum_{l = 1}^{n_{g}}
    \left(
      \bar{Y}_{l}-Y^{t}_{l}
    \right)^{2}
  }.
  \label{eq:rmse}
\end{equation}
Here, $n_{g}$ is the number of grid cells,
$\bar{\mathbf{Y}} \in R^{n_{g}}$ is the vector containing the
\emph{estimated} mean logarithmic permeabilities across the model
domain and $\mathbf{Y}^{t} \in R^{n_{g}}$ is the vector containing the
corresponding \emph{synthetic reference} logarithmic permeabilities. %

For each EnKF-method and for a given ensemble size, a large number of
synthetic experiments is computed. %
These synthetic experiments differ solely in the perturbation of
initial fields and measurements. %
A single synthetic experiment provides a sample-RMSE, calculated
according to Equation (\ref{eq:rmse}) and called
$r_{i}^{a,\,n_{e}}$. %
Taken together, the $n_{syn} = 1000$ synthetic experiments (or in the
case of large ensembles $n_{syn} = 100$ synthetic experiments) for a
given method $a$ and ensemble size $n_{e}$ provide an approximate
probability density function of the RMSE. %
We calculate RMSE means according to: %
\begin{equation}
  \bar{r}^{a,\,n_{e}}  =
  \frac{1}{n_{syn}}\sum_{i=1}^{n_{syn}}r_{i}^{a,\,n_{e}}.
  \label{eq:rmse-mean}
\end{equation}
We use these RMSE means to compare EnKF-methods. %

The question arises, whether a small number of synthetic experiments
$n_{syn}$ would suffice to evaluate the performance of the
EnKF-methods. %
We compare RMSE means of two EnKF-methods on the basis of
$n_{syn}= 1,\, 10,$ or $100$ synthetic experiments. %
Ten thousand subsets $X$ of $n_{syn}$ synthetic experiments are
randomly sampled from the $1000$ synthetic experiments. %
We calculate and compare the corresponding means
$\bar{r}^{a,\,n_{e}}_{X}$. %
The fraction, for which one EnKF-method $a$ yields a smaller RMSE than
another EnKF-method $b$ %
\begin{equation}
  p^{a<b,\,n_{e}} =
  \frac{\# \{
    X \mid \bar{r}^{a,\,n_{e}}_{X} < \bar{r}^{b,\,n_{e}}_{X}
    \}
  }{10000},
  \label{eq:rmse-fraction}
\end{equation}
is recorded (the sum $p^{a<b,\,n_{e}}+p^{b<a,\,n_{e}}$ is by
definition equal to 1.0). %
Doing so, we estimate the probability that one method outperforms
another method on the basis of $n_{syn}$ synthetic experiments. %

Quotients of the RMSE means (based on all 1000 synthetic experiments)
for all combinations of EnKF-variants are calculated: %
\begin{equation}
  q^{a<b,\,n_{e}} = \frac{\bar{r}^{a,\,n_{e}}}{\bar{r}^{b,\,n_{e}}}.
  \label{eq:quots}
\end{equation}
In this formula we choose EnKF-method $a$ to have the smaller RMSE
mean, so that $q^{a<b,\,n_{e}} \leq 1$ . %
From the quotients, we calculate relative differences: %
\begin{equation}
  d^{a<b,\,n_{e}} =
  \frac{\bar{r}^{b,\,n_{e}}-\bar{r}^{a,\,n_{e}}}{\bar{r}^{b,\,n_{e}}}
  = 1-q^{a<b,\,n_{e}}.
  \label{eq:reldiffs}
\end{equation}
For example, it could be that EnKF-variant $a$ gives on average
(calculated over 1000 synthetic experiments) a RMSE which is
$d^{a<b,\,n_{e}}=10\%$ smaller than another EnKF-variant $b$. %
In that case, the quotient would be $q^{a<b,\,n_{e}}=0.9$. %
We analyze which relative differences are with at least 95\%
probability statistically significant for $n_{syn}$ synthetic
experiments and which are not. %
\begin{equation}
  D^{n_{e}}_{+} = \{ d^{a<b,\,n_{e}} \mid p^{a<b,\,n_{e}} > 0.95 \}
  \label{eq:sig-reldiffs}
\end{equation}
\begin{equation}
  D^{n_{e}}_{-} = \{ d^{a<b,\,n_{e}} \mid p^{a<b,\,n_{e}} \leq 0.95 \}
  \label{eq:insig-reldiffs}
\end{equation}
Finally, we compare the smallest significant relative difference %
\begin{equation}
  d^{min}_{+} = \min D^{n_{e}}_{+}
  \label{eq:min-sig}
\end{equation}
to the largest insignificant relative difference %
\begin{equation}
  d^{max}_{-} = \max D^{n_{e}}_{-}.
  \label{eq:max-insig}
\end{equation}
In general, one would expect $d^{min}_{+}$ to be larger than
$d^{max}_{-}$. %
In this case, we choose the threshold for significant relative
differences between the two values. %
However, due to the specific shape of the RMSE distributions
(especially related to their spread), it may occur that
$d^{min}_{+}<d^{max}_{-}$. %
If this happens, we evaluate the comparisons on the basis of their
specific distributions and choose the threshold manually. %

\section{Results and Discussion}
\label{sec:discussion-results}

\subsection{Tracer model}
\label{sec:results-tracer}

\subsubsection{Comparison of EnKF-variants in terms of Mean RMSE}
\label{sec:mean-rmse-tracer}

\begin{figure}
  \centering
  \includegraphics[trim = 50 170 50 130, clip,
  width=\columnwidth]{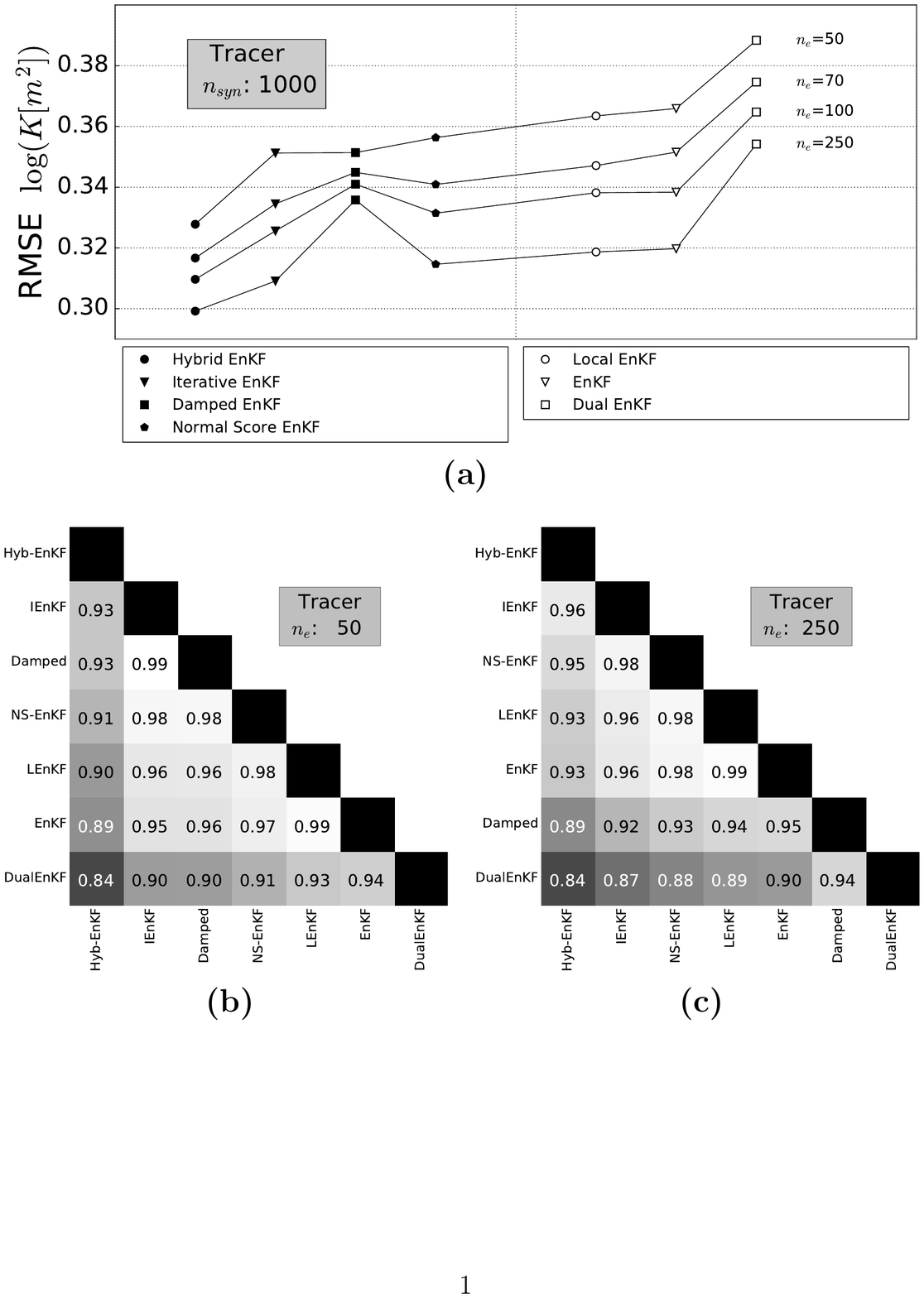}
  \caption{Mean RMSEs for seven EnKF-methods derived from 1000
    synthetic experiments for the tracer model - Ensemble Sizes:
    $n_{e} = 50,\, 70,\, 100,\, 250$ (a). For ensemble sizes
    $n_{e}=50 \textrm{ (b)},\, 250 \textrm{ (c)}$, the RMSE mean for a
    method on the horizontal axis divided by the RMSE mean for a
    method on the vertical axis is given.}
  \label{fig:wavereal-errorplot}
\end{figure} %

Figure \ref{fig:wavereal-errorplot} shows the mean RMSEs for all
methods and ensemble sizes 50, 70, 100 and 250. %
The RMSE values $\bar{r}^{a,\,n_{e}}$ range from
$0.3\, \log_{10}(K[\mathrm{m}^{2}])$ to
$0.4\, \log_{10}(K[\mathrm{m}^{2}])$. %
For ensemble size 50, Hybrid EnKF yields the smallest mean RMSE
($0.328\, \log_{10}(K[\mathrm{m}^{2}])$ followed by Iterative EnKF,
Damped EnKF and Normal Score EnKF
($0.351\, \log_{10}(K[\mathrm{m}^{2}])$,
$0.351\, \log_{10}(K[\mathrm{m}^{2}])$ and
$0.356\, \log_{10}(K[\mathrm{m}^{2}])$). %
Local EnKF ($0.363\, \log_{10}(K[\mathrm{m}^{2}])$) and Classical EnKF
($0.366\, \log_{10}(K[\mathrm{m}^{2}])$) follow. %
Finally, Dual EnKF yields the largest RMSE:
$0.388\, \log_{10}(K[\mathrm{m}^{2}])$. %

When the ensemble size is increased from 50 to 250, all methods
produce ensemble mean permeability fields closer to the reference. %
The RMSE for Damped EnKF is
$\bar{r}^{\mathrm{Damped},\,50}-\bar{r}^{\mathrm{Damped},\,250}=0.016\,
\log_{10}(K[\mathrm{m}^{2}])$ smaller for an ensemble size of 250 than
for an ensemble size of 50. %
This RMSE reduction is the smallest of all EnKF-methods and the next
smallest reduction
($\bar{r}^{\mathrm{Hyb-EnKF},\,50}-\bar{r}^{\mathrm{Hyb-EnKF},\,250}=0.029\,
\log_{10}(K[\mathrm{m}^{2}])$ smaller for ensemble size 250 than for
ensemble size 50) is by Hybrid EnKF. %
As a result, Damped EnKF has the second largest RMSE among all
EnKF-methods for ensemble size 250. %
The ranking of the remaining methods is identical for all four
ensemble sizes. %

The second part of Figure \ref{fig:wavereal-errorplot} contains tables
of quotients $q^{a<b,\,50}$ and $q^{a<b,\,250}$ (as defined in
Equation (\ref{eq:quots})) for all pairs of EnKF-methods. %
The largest relative difference in the tables is $16\%$
(\textit{Hybrid EnKF vs Dual EnKF}) for both ensemble sizes 50 and
250. %
Three relative differences are as small as $1\%$ (\textit{Iterative
  EnKF vs Damped EnKF} and \textit{Local EnKF vs Classical EnKF} for
ensemble size 50, \textit{Local EnKF vs Classical EnKF} for ensemble
size 250). %
In the following, we examine which relative differences can be
considered significant as function of the number of synthetic
experiments. %

It is worth mentioning that the mean RMSE difference between initial
permeability fields and the synthetic truth
($0.62\, \log_{10}(K[\mathrm{m}^{2}])$) is significantly reduced in
all synthetic experiments. %
Finally, we note that the large RMSEs of Local EnKF, Classical EnKF
and Dual EnKF can be attributed at least partially to a non-negligible
number of synthetic experiments with very large RMSEs. %
The full RMSE distributions are shown in Figure
\ref{fig:rmse-distributions}. %

\subsubsection{Thresholds on significant RMSE differences}
\label{sec:mean-rmse-tracer-less-syn-exp}

\begin{figure}
  \centering
  \includegraphics[trim = 50 50 50 50, clip,
  width=\columnwidth]{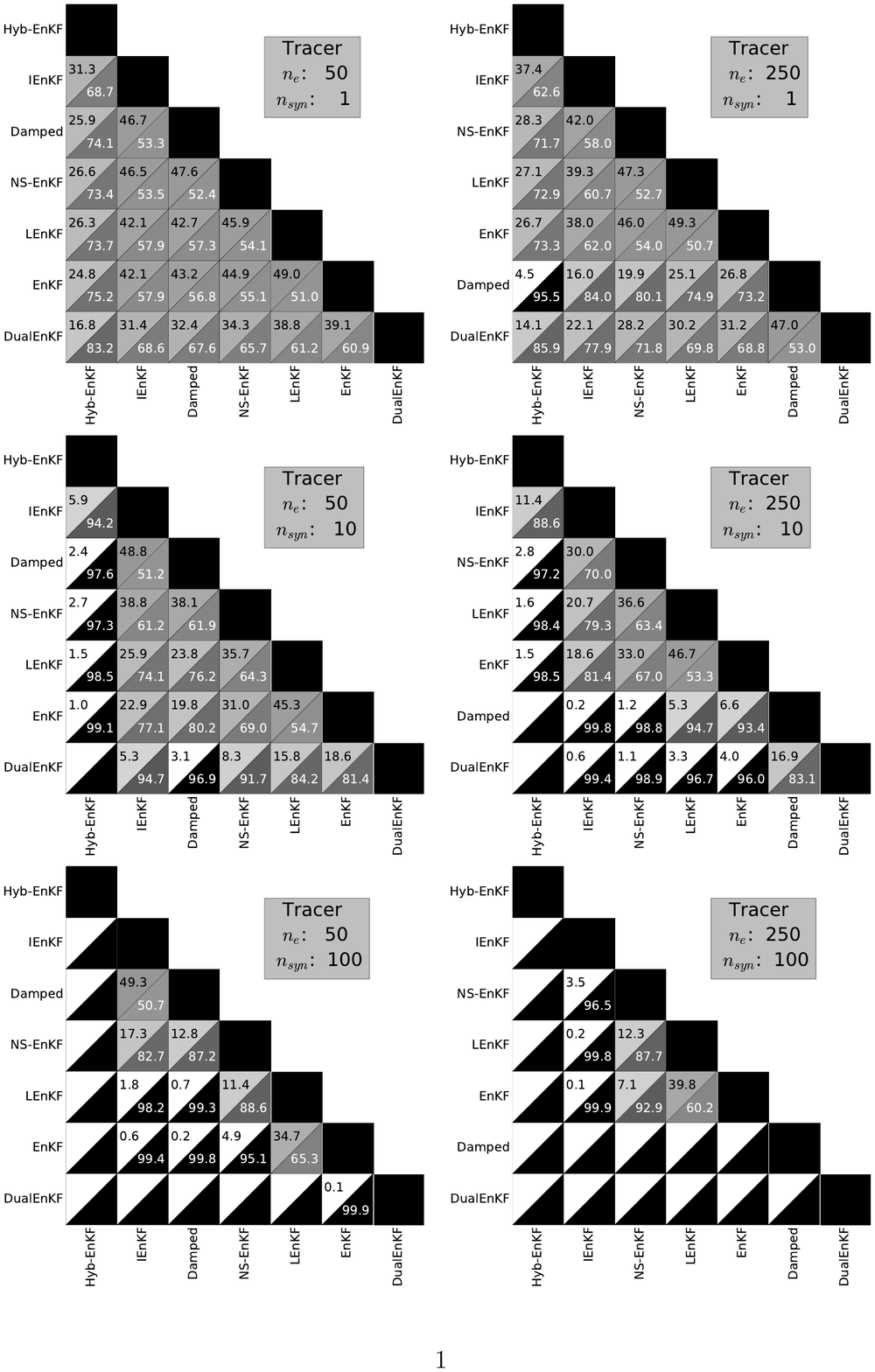}
  \caption{Probabilities to find a smaller RMSE mean for a given
    EnKF-variant compared to another EnKF-variant, based on 1, 10 and
    100 synthetic experiments (tracer model, ensemble sizes 50 and
    250).}
  \label{fig:wavereal-numcomp}
\end{figure}

It is evaluated whether significant differences in performance between
EnKF-variants can be demonstrated for 1, 10 or 100 synthetic
experiments. %
Figure \ref{fig:wavereal-numcomp} shows probabilities to find a
smaller RMSE mean for a given EnKF-variant compared to another
EnKF-variant based on $n_{syn}$ synthetic experiments. %

The probabilities $p^{a<b,\,50}$ and $p^{a<b,\,250}$ from Figure
\ref{fig:wavereal-numcomp} are cross-checked with the quotients
$q^{a<b,\,50}$ and $q^{a<b,\,250}$ in Figure
\ref{fig:wavereal-errorplot} to determine thresholds for significant
relative RMSE differences. %
Significant relative differences are defined as the relative
differences, for which at least $95\%$ of the comparisons favor one of
the two EnKF-methods. %
Based on a single synthetic experiment, and for ensemble size 50, all
relative RMSE differences are insignificant. %
Consequently, RMSE differences between two EnKF-methods smaller than
$16\%$ cannot be detected on the basis of a single synthetic
experiment and ensemble size 50. %
For ensemble size 250, the RMSE difference of $11\%$ for the
comparison pair \textit{Hybrid EnKF vs Damped EnKF} is significant. %
On the other hand, there are three insignificant comparisons with
larger relative differences (\textit{Hybrid EnKF vs Dual EnKF},
\textit{Iterative EnKF vs Dual EnKF} and \textit{Normal Score EnKF vs
  Dual EnKF}). %
The anomaly that three insignificant relative differences are larger
than a significant relative difference is due to the narrow RMSE
distribution of the Damped EnKF (Figure \ref{fig:rmse-distributions}
shows that there are fewer RMSEs below
$0.3 \, \log_{10}(K[\mathrm{m}^{2}])$ for Damped EnKF than for any
other method). %
Taking this into account, we set the threshold for which significant
RMSE differences can be detected to $15\%$. %
For 10 synthetic experiments and ensemble size 50, there are RMSE
differences of $7\%$ and $9\%$ that are significant (\textit{Hybrid
  EnKF vs Damped EnKF}, \textit{Hybrid EnKF vs Normal Score EnKF}) and
differences of $9\%$ and $10\%$ that are not (\textit{Normal Score
  EnKF vs Dual EnKF}, \textit{Iterative EnKF vs Dual EnKF}). %
Taking into account the narrow RMSE distribution of the Damped EnKF,
the threshold is set to $9\%$. %
For ensemble size 250, and 10 synthetic experiments, there is a RMSE
difference of $5\%$ that is significant (\textit{Hybrid EnKF vs Normal
  Score EnKF}) and there are two of $6\%$ that are not (\textit{Hybrid
  EnKF vs Normal Score EnKF}, \textit{Damped EnKF vs Dual EnKF}). %
The threshold is set to $6\%$. %
For 100 synthetic experiments and ensemble size 50, there is a
difference of $3\%$ that is significant (\textit{Normal Score EnKF vs
  Classical EnKF}) and there are three differences of $2\%$ that are
not (\textit{Iterative EnKF vs Normal Score EnKF}, \textit{Damped EnKF
  vs Normal Score EnKF}, \textit{Normal Score EnKF vs Local EnKF}). %
The threshold is set to $2\%$. %
For ensemble size 250, there is a difference of $2\%$ that is
significant (\textit{Iterative EnKF vs Normal Score EnKF}) and there
are two differences of $2\%$ that are not (\textit{Normal Score EnKF
  vs Local EnKF}, \textit{Normal Score EnKF vs Classical EnKF}). %
The threshold is set to $2\%$. %
All thresholds are provided in Table \ref{tab:numcomp-summary}. %

\subsection{Well model}
\label{sec:results-well}

\subsubsection{Comparison of EnKF-variants in terms of Mean RMSE}
\label{sec:mean-rmse-well}

For the well model, EnKF-variants (and ensemble sizes) are compared on
the basis of the calculated RMSE means in the same way as for the
tracer model. %
Results for ensemble sizes 50, 70, 100 and 250 are discussed here. %
Results for ensemble sizes 500, 1000 and 2000 are discussed later in
Section \ref{sec:large-ensembles}. %

\begin{figure}
  \centering
  \includegraphics[trim = 50 170 50 130, clip,
  width=\columnwidth]{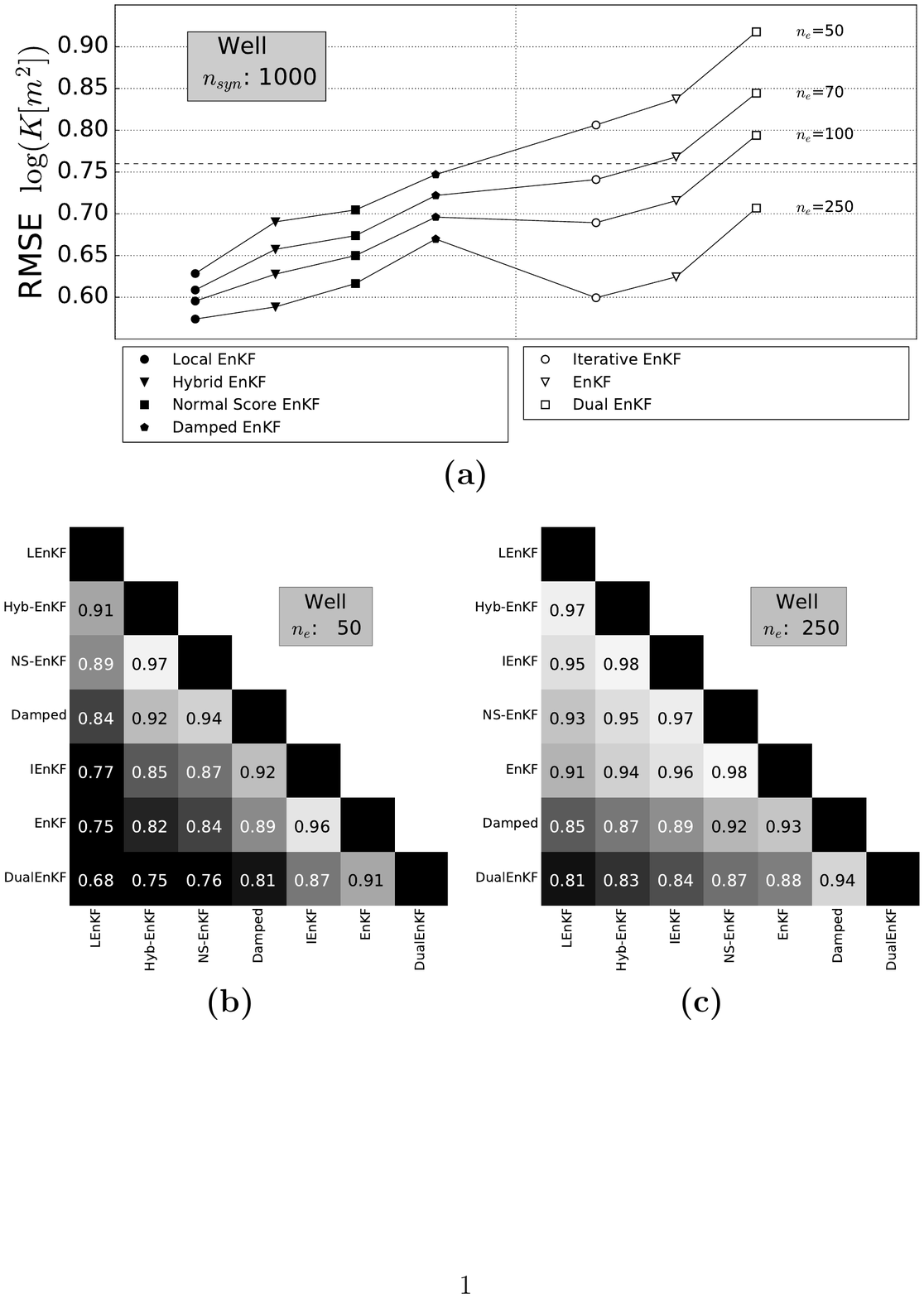}
  \caption{Mean RMSEs for seven EnKF-methods derived from 1000
    synthetic experiments for the well model - Ensemble Sizes:
    $n_{e} = 50,\, 70,\, 100,\, 250$ (a). For ensemble sizes
    $n_{e}=50, \textrm{ (b)},\, 250 \textrm{ (c)}$, the RMSE mean for
    a method on the horizontal axis divided by the RMSE mean for a
    method on the vertical axis is given.}
  \label{fig:wavewell-errorplot}
\end{figure} %

Figure \ref{fig:wavewell-errorplot} shows the mean RMSEs for all
methods and ensemble sizes. %
The values range from $0.55\, \log_{10}(K[\mathrm{m}^{2}])$ to
$0.95\, \log_{10}(K[\mathrm{m}^{2}])$. %
For the smallest ensemble size of 50, observed mean RMSEs show the
largest spread. %
Local EnKF yields the smallest mean RMSE
($0.63\, \log_{10}(K[\mathrm{m}^{2}])$ followed by Hybrid EnKF, Normal
Score EnKF and Damped EnKF ($0.69\, \log_{10}(K[\mathrm{m}^{2}])$,
$0.70\, \log_{10}(K[\mathrm{m}^{2}])$ and
$0.75\, \log_{10}(K[\mathrm{m}^{2}])$). %
The methods Iterative EnKF, Classical EnKF and Dual EnKF
($0.81\, \log_{10}(K[\mathrm{m}^{2}])$ and
$0.84\, \log_{10}(K[\mathrm{m}^{2}])$,
$0.92\, \log_{10}(K[\mathrm{m}^{2}])$) yield mean RMSEs that are
larger than those for the initial permeability fields suggesting
divergence of the algorithm for a significant fraction of the
synthetic experiments. %

When the ensemble size is increased from 50 to 250, all methods get
closer to the reference. %
For the methods Local EnKF, Hybrid EnKF, Normal Score EnKF and Damped
EnKF the RMSE reduces by up to $0.1\, \log_{10}(K[\mathrm{m}^{2}])$,
when the ensemble size is increased from 50 to 250. %
For Iterative EnKF, Classical EnKF and Dual EnKF this reduction is
around $0.2\, \log_{10}(K[\mathrm{m}^{2}])$. %
As a result, Iterative EnKF ends up with the third smallest RMSE for
ensemble size 250, Classical EnKF performs better than Damped EnKF and
Dual EnKF still has the largest RMSE, but much closer to the other
methods. %

The second part of Figure \ref{fig:wavewell-errorplot} contains tables
of quotients $q^{a<b,\,50}$ and $q^{a<b,\,250}$ (as defined in
Equation (\ref{eq:quots})) for all pairs of EnKF-methods. %
For the well model, the relative differences are much larger than for
the tracer model. %
The largest relative differences in the tables are $32\%$ for ensemble
size 50 and $19\%$ for ensemble size 250 (both \textit{Local EnKF vs
  Dual EnKF}). %
For ensemble size 50, the smallest relative difference is $3\%$ for
comparison pair \textit{Hybrid EnKF vs Normal Score EnKF}. %
For ensemble size 250, two relative differences are at $2\%$
(\textit{Hybrid EnKF vs Iterative EnKF}, \textit{Normal Score EnKF vs
  Classical EnKF}). %
In general, relative differences are smaller for the larger ensemble
size. %

The much lower RMSE for ensemble size 250 (compared to ensemble size
50) is most probably related to the overall worse performance of the
data assimilation experiments for the well model compared to the
tracer model. %
The worse performance is related to the much smaller correlation
length, in combination with the high variance of the permeability
field, for this set-up. %
The benefit of estimating noisy model covariances with a larger
ensemble seems therefore to be higher than for the tracer model. %

\subsubsection{Thresholds on significant RMSE differences}
\label{sec:mean-rmse-well-less-syn-exp}

Also for the well model, it is evaluated whether significant
differences in performance between EnKF-variants can be demonstrated
with 1, 10 or 100 synthetic experiments. %

\begin{figure}
  \centering
  \includegraphics[trim = 50 50 50 50, clip,
  width=\columnwidth]{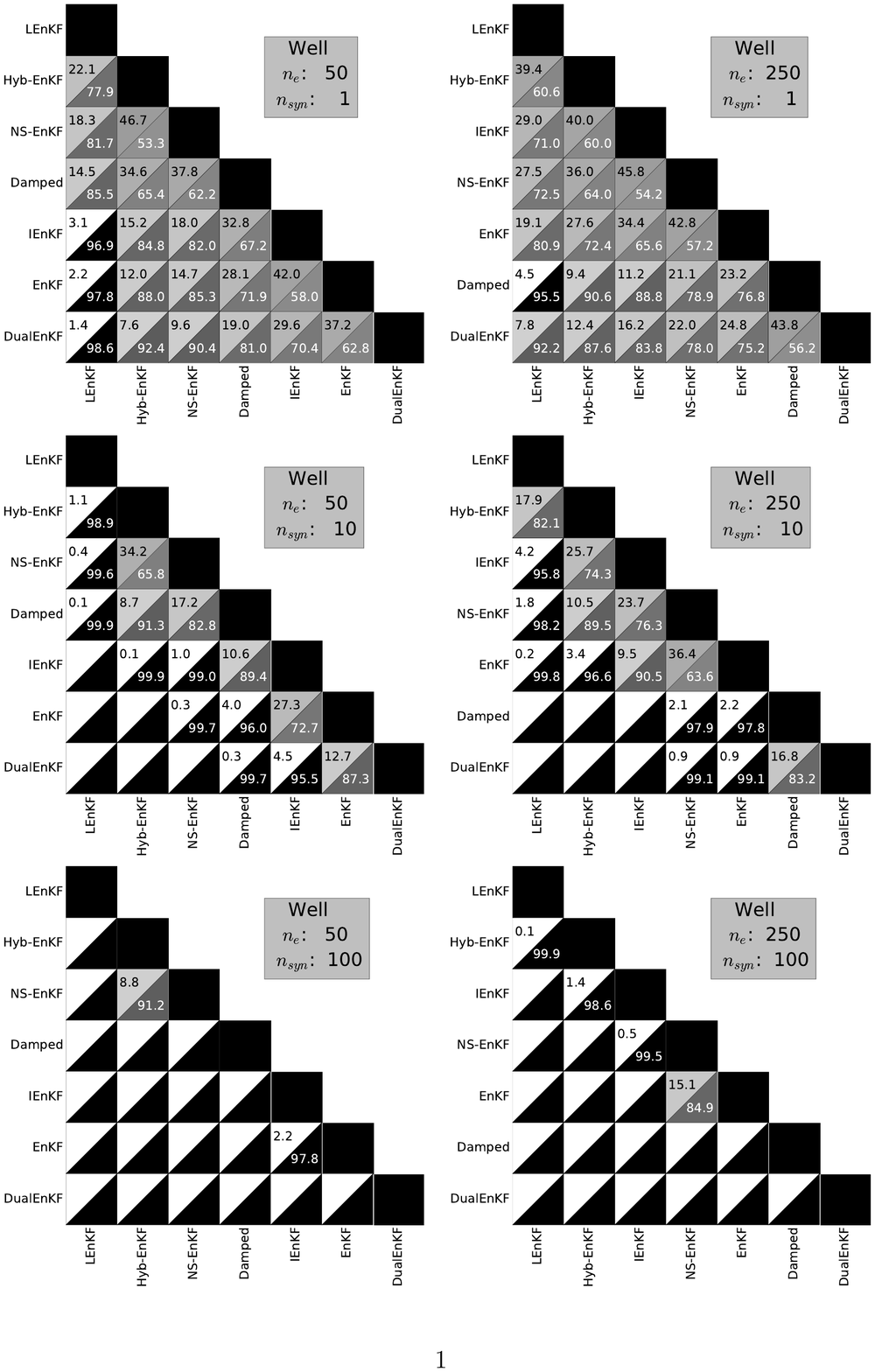}
  \caption{The same as Figure \ref{fig:wavereal-numcomp} for the well
    model.}
  \label{fig:wavewell-numcomp}
\end{figure}%

The probabilities $p^{a<b,\,50}$ and $p^{a<b,\,250}$ from Figure
\ref{fig:wavewell-numcomp} are cross-checked with the quotients
$q^{a<b,\,50}$ and $q^{a<b,\,250}$ in Figure
\ref{fig:wavewell-errorplot} to determine thresholds for significant
relative RMSE differences. %
For a single synthetic experiment and ensemble size 50, there are
significant RMSE differences of $25\%$ and $23\%$ (\textit{Local EnKF
  vs Classical EnKF}, \textit{Local EnKF vs Iterative EnKF}) and
insignificant RMSE differences of $24\%$ and $25\%$ (\textit{Normal
  Score EnKF vs Dual EnKF}, \textit{Hybrid EnKF vs Dual EnKF}). %
The threshold is set to $25\%$. %
For ensemble size 250, there is the significant case \textit{Local
  EnKF vs Damped EnKF} for $15\%$ difference in mean RMSE, but there
are also three comparisons with differences $16\%$, $17\%$ and $19\%$
that are not significant (\textit{Iterative EnKF vs Dual EnKF},
\textit{Hybrid EnKF vs Dual EnKF}, \textit{Local EnKF vs Dual
  EnKF}). %
As for the tracer set-up, the smallest significant RMSE difference is
explained by the relatively narrow distribution of RMSEs for Damped
EnKF (compare Figure \ref{fig:rmse-distributions}). %
The threshold is set to $18\%$. %
For 10 synthetic experiments and ensemble size 50, there are two
differences of $9\%$, one is significant (\textit{Local EnKF vs Hybrid
  EnKF}) and one is not (\textit{Classical EnKF vs Dual EnKF}). %
Thus, the threshold is set to $9\%$. %
For ensemble size 250, there is a difference of $5\%$ that is
significant (\textit{Local EnKF vs Iterative EnKF}) and there is one
of $6\%$ that is not (\textit{Damped EnKF vs Dual EnKF}). %
The threshold is set to $5\%$. %
For 100 synthetic experiments and ensemble size 50, there is a
difference of $4\%$ that is significant (\textit{Iterative EnKF vs
  Classical EnKF}) and there is one of $3\%$ that is not
(\textit{Hybrid EnKF vs Normal Score EnKF}). %
The threshold is set to $3\%$. %
For ensemble size 250, there is a difference of $2\%$ that is
significant (\textit{Hybrid EnKF vs Iterative EnKF}) and there is
another one of $2\%$ that is not (\textit{Normal Score EnKF vs
  Classical EnKF}). %
The threshold is set to $2\%$. %
All thresholds are provided in Table \ref{tab:numcomp-summary}. %

\begin{table}
  \centering
  \caption{RMSE differences which are considered to be significant for
    the two simulation set-ups and as a function of the number of
    synthetic experiments $n_{syn}$ and the number of ensemble members
    $n_{e}$.} %
  \label{tab:numcomp-summary}
  \begin{tabular}{l l c c c c c c c}
    \hline

    &
      $n_{syn}$
    &
      $n_{e} = 50$
    &
      $n_{e} = 70$
    &
      $n_{e} = 100$
    &
      $n_{e} = 250$
    &
      $n_{e} = 500$
    &
      $n_{e} = 1000$
    &
      $n_{e} = 2000$
    \\
    \hline                      
    Tracer
    &
      1
    &
      $>\,16\%$
    &
      $>\,16\%$
    &
      $>\,16\%$
    &
      15\%
    &
      13\%
    &
      15\%
    &
      14\%
    \\

    &
      10
    &
      9\%
    &
      9\%
    &
      8\%
    &
      6\%
    &
      6\%
    &
      5\%
    &
      4\%
    \\

    &
      100
    &
      2\%
    &
      3\%
    &
      3\%
    &
      2\%
    &
      -
    &
      -
    &
      -
    \\
    \hline                      
    Well
    &
      1
    &
      25\%
    &
      20\%
    &
      23\%
    &
      18\%
    &
      17\%
    &
      16\%
    &
      15\%
    \\

    &
      10
    &
      9\%
    &
      9\%
    &
      8\%
    &
      5\%
    &
      5\%
    &
      5\%
    &
      5\%
    \\

    &
      100
    &
      3\%
    &
      $<\, 3\%$
    &
      2\%
    &
      2\%
    &
      -
    &
      -
    &
      -
    \\
    \hline
  \end{tabular}
\end{table} %

\subsection{Large ensembles}
\label{sec:large-ensembles}

\begin{figure}
  \centering
  \includegraphics[trim =  50 200 50 130, clip,
  width=\columnwidth]{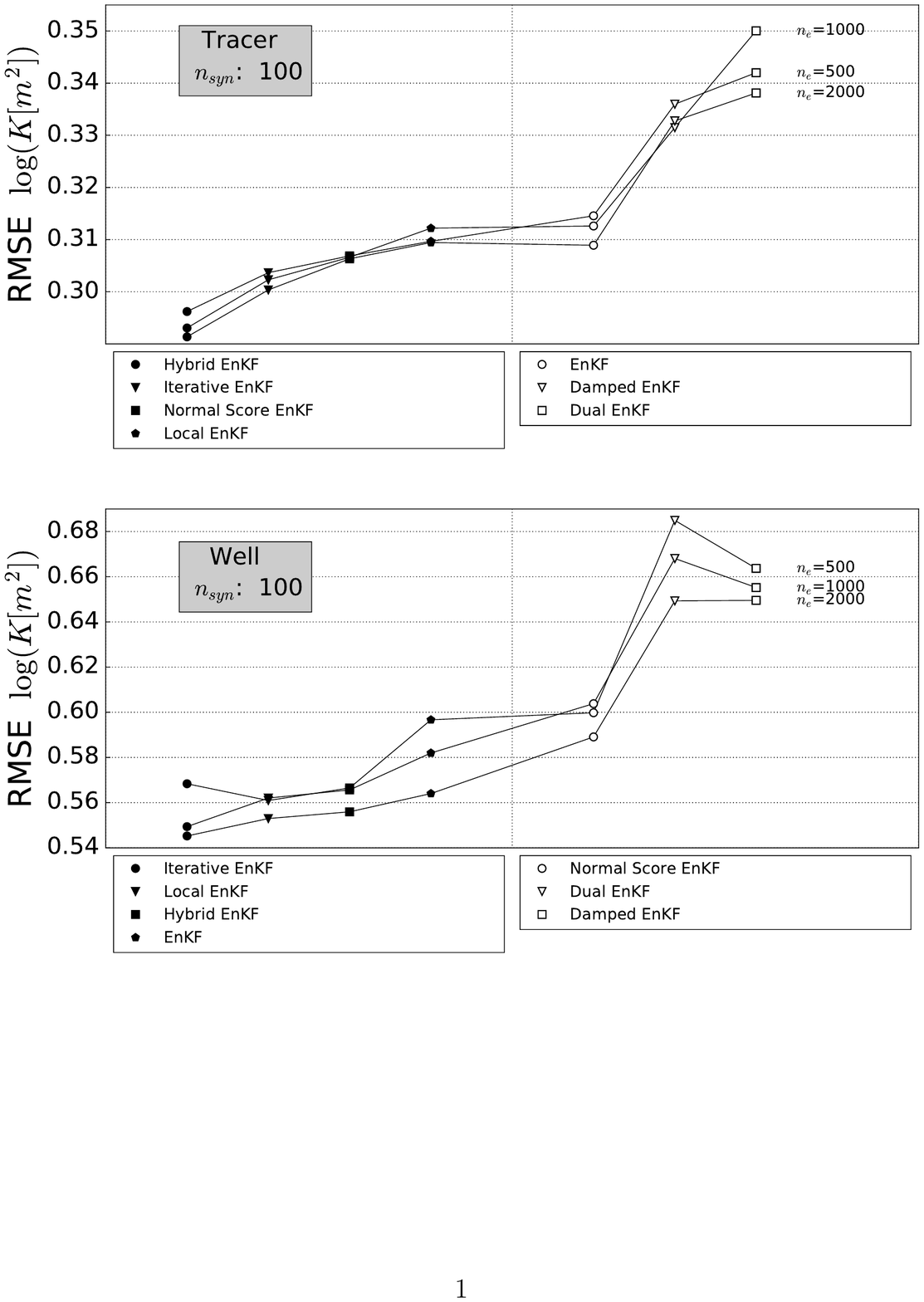}
  \caption{Mean RMSEs for seven EnKF-methods derived from 100
    synthetic experiments for the tracer and well model and ensemble
    sizes 500, 1000 and 2000.}
  \label{fig:large-ensembles-well-tracer}
\end{figure} %

Results from synthetic experiments for ensemble sizes 500, 1000 and
2000 are shown in Figure \ref{fig:large-ensembles-well-tracer}. %
RMSE means are computed from 100 synthetic experiments instead of the
1000 synthetic experiments which were calculated for the smaller
ensemble sizes. %
This leads to a larger uncertainty in the mean calculation. %
Additionally, for ensemble sizes 500, 1000 and 2000, RMSE means are
very similar. %
RMSE means are even not always smaller for a larger ensemble size (for
example for Dual EnKF and the tracer model, the RMSE mean for ensemble
size 500 is smaller than the RMSE mean for ensemble size 1000). %
The ranking of the performance of the different EnKF-variants for the
large ensemble sizes is similar to the method ranking for ensemble
size 250. %
In addition, for larger ensemble sizes the number of synthetic
experiments which is needed to show significant difference in
performance of EnKF-methods is only marginally smaller than for
ensemble size 250 (see Table \ref{tab:numcomp-summary}). %

\subsection{RMSE distributions}
\label{sec:rmse-distributions}

\begin{figure}
  \centering
  \includegraphics[trim =  50 150 50 120, clip,
  width=\columnwidth]{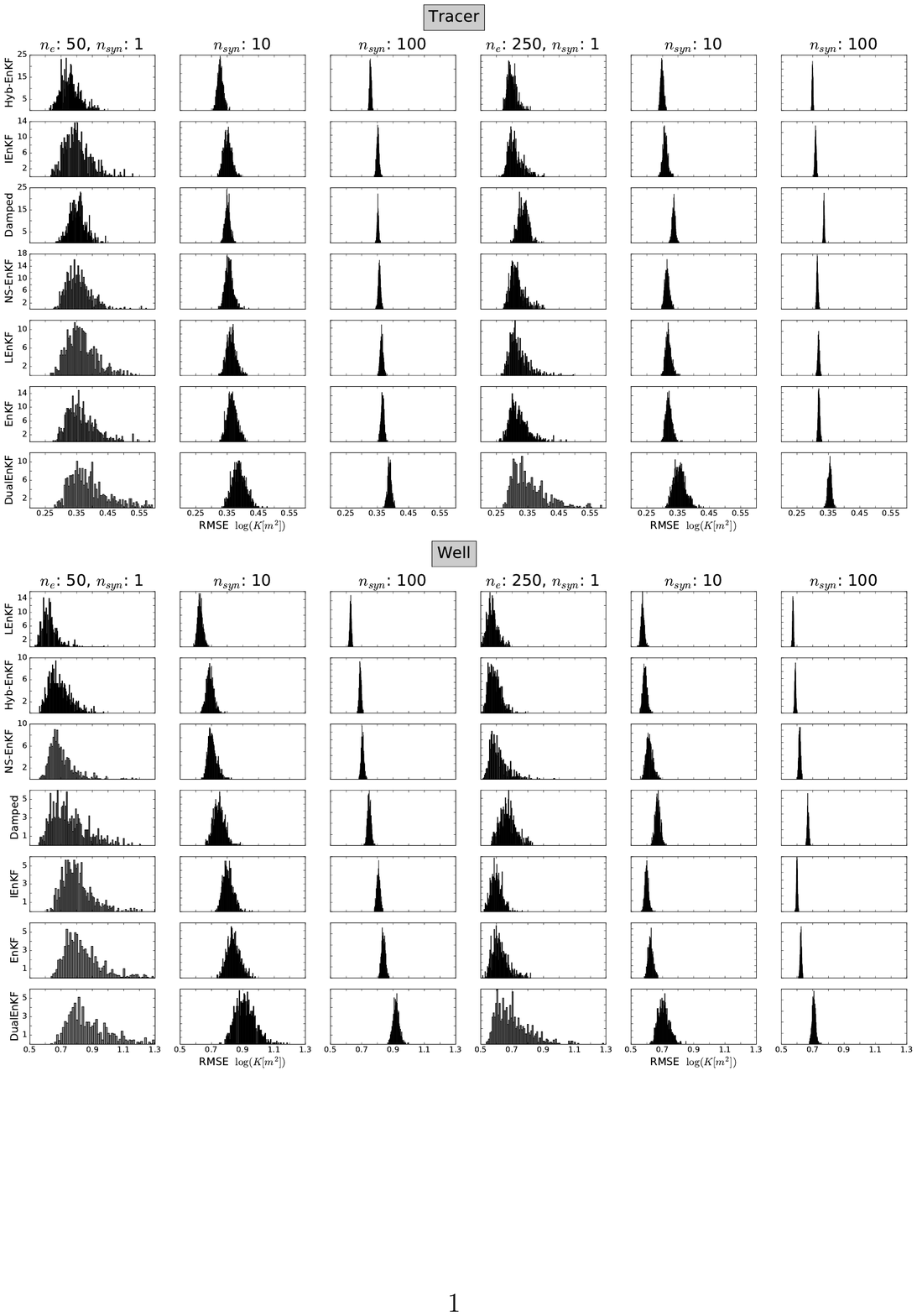}
  \caption{Mean RMSE distributions for ensemble sizes 50 and 250, for
    1, 10 and 100 synthetic experiments, for the tracer and well
    model.}
  \label{fig:rmse-distributions}
\end{figure} %

The mean RMSE distributions calculated for the two model set-ups,
$n_{syn} = 1, 10$ or $100$ and ensemble sizes 50 and 250 are displayed
in Figure \ref{fig:rmse-distributions}. %
For $n_{syn} = 10$ or $100$, the mean RMSE distributions are close to
Gaussian, but for $n_{syn} = 1$ the spread is larger with more
outliers. %
The narrow distributions calculated on the basis of 100 synthetic
experiments, for the different EnKF-methods, show little overlap. %
Comparing RMSE distributions for ensemble size 50 and 250, it can be
seen that there are less outliers for the larger ensemble size. %
If we compare the RMSE distributions for the tracer and well model, it
is clear that especially Damped EnKF has a different distribution from
the rest. %
Whereas it has the narrowest distributions for the tracer set-up, it
has one of the widest distributions for the well set-up. %

The width of the RMSE distributions illustrates the variability of the
estimation results related to the random seed variation. %
The large overlap of RMSE distributions for the different
EnKF-variants suggests that the estimation result is more dependent on
the random seed than on the choice of EnKF-method. %
If we base the analysis on 10 or 100 synthetic experiments, the
distributions get narrower allowing to detect significant differences
in performance of different EnKF-methods. %

\subsection{Variation of EnKF-method parameters}
\label{sec:vari-enkf-method-params}

\begin{figure}
  \centering
  \includegraphics[trim = 50 270 50 230, clip,
  width=\columnwidth]{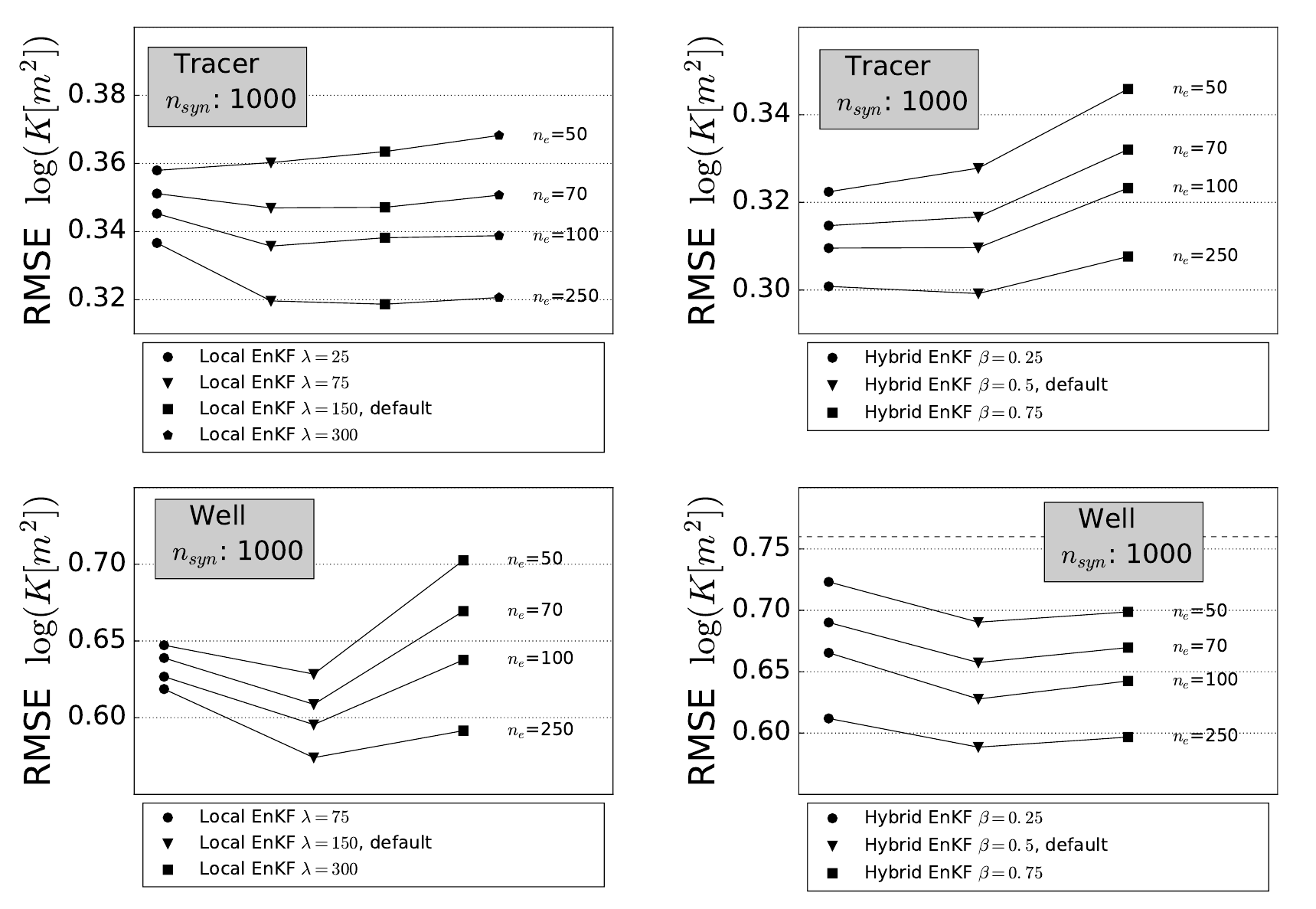}
  \caption{Variation of mean RMSE for Local EnKF and Hybrid EnKF, as a
    function of values for the parameters $\lambda$ and $\beta$ and
    ensemble size. Results are shown for both the tracer model and the
    well model.}
  \label{fig:enkf-method-params}
\end{figure} %

For the two methods Local EnKF and Hybrid EnKF, the correlation length
$\lambda$ and mixing parameter $\beta$ were varied, respectively. %
The resulting mean RMSEs are displayed in Figure
\ref{fig:enkf-method-params}. %
For the tracer set-up and ensemble size 50, Local EnKF with the
smallest correlation length of $25\, \mathrm{m}$ yields the smallest
mean RMSE. %
On the contrary, for larger ensemble sizes the smallest correlation
length yields the largest mean RMSE and a correlation length of
$150\, \mathrm{m}$ results in the smallest mean RMSE. %
For the well model, the correlation length of $150\, \mathrm{m}$
yields the smallest mean RMSEs for all ensemble sizes. %
Although the results are affected by the parameter values, the
performance of Local EnKF is not so strongly influenced by the choice
of the correlation length in these cases, except for ensemble size
250, for which a small correlation length results in a large RMSE
affecting the ranking of the EnKF methods. %

For Hybrid EnKF, the parameter $\beta$ also has a certain impact on
the performance which is larger for small ensemble sizes. %
This influences the ranking of the methods, but to a limited extent. %
These examples also show the importance of parameter settings which
add an additional uncertainty component to model comparisons. %

\subsection{Varying the observation noise}
\label{sec:vary-obs-noise}

\begin{figure}
  \centering
  \includegraphics[trim =  50 350 50 300, clip,
  width=\columnwidth]{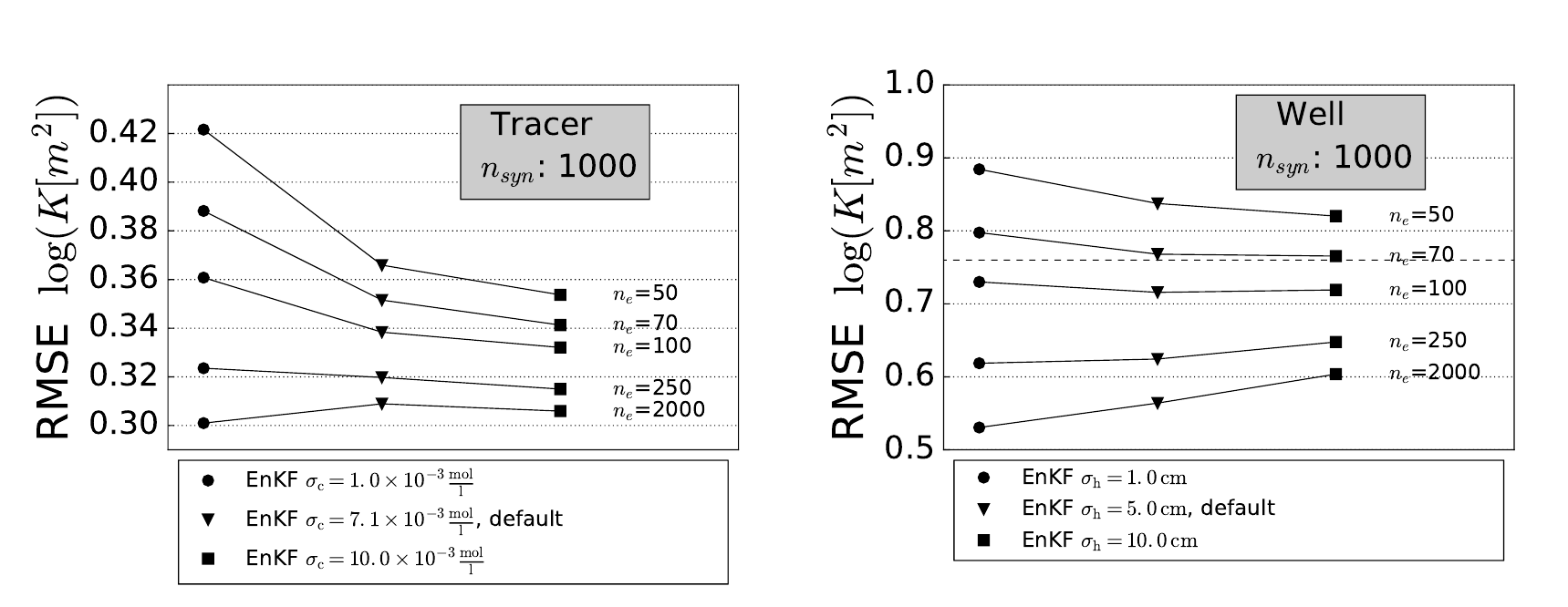}
  \caption{Variation of mean RMSE for different observation noises in
    Classical EnKF, as function of ensemble size. Results are shown
    for both the tracer model and the well model.}
  \label{fig:vary-obs-noise}
\end{figure} %

To illustrate the effect of different observation noises, the tracer
and well set-ups were computed for Classical EnKF and observation
noises larger and smaller than the noise level chosen for the default
synthetic experiment. %
In Figure \ref{fig:vary-obs-noise}, the resulting RMSE means are shown
for ensemble sizes 50, 70, 100, 250 and 2000. %
For ensemble size 50, in both set-ups, smaller observation noise leads
to larger RMSE means, whereas the larger observation noise leads to
smaller RMSE means. %
This trend changes, when the ensemble size is larger. %
For ensemble size 2000, in both set-ups, the smallest observation
noise results in the smallest RMSE mean. %

Results for different observation noises can be understood taking into
account two different impacts of observation noise on the simulation
results. %
First, smaller observation noise results in a higher weight for
observations, i.e. a stronger correcting influence of the
observations. %
It should result on average in smaller RMSEs as long as its influence
is correctly weighted in the filter and it is therefore important that
model covariances are correct, which is the case if the ensemble size
is large. %
For small ensemble sizes, model covariances differ more strongly from
the true values resulting in an incorrect weighting and the risk that
updates go in the wrong direction resulting in larger RMSEs. %
The default values were originally chosen because they represent the
behavior of typical measurement devices. %
According to these results, they also seem to constitute a sensible
trade-off between filter divergence due to unjustifiably small
observation noise and information loss due to unjustifiably large
observation noise. %

\subsection{Discussion}
\label{sec:add-discussion}

The tracer model and the well model have different boundary
conditions, different correlation lengths of the reference
permeability fields and different measurement types and amounts which
are assimilated. %
The data assimilation experiments for each of these two model set-ups
result in different RMSE values. %
Nevertheless, for the two model set-ups similar conclusions are
reached in terms of the number of synthetic experiments which is
needed to show that one EnKF-variant outperforms another one. %
Given these results, we recommend that for comparisons of
EnKF-variants 10 or more synthetic experiments are needed. %
A number of 10 synthetic experiments is needed to show that one
EnKF-variant significantly outperforms another EnKF-variant if the two
methods show a difference in RMSE of at least 10\%. %
It can be argued that a RMSE difference of 10\% is important enough to
be detected. %
For really small RMSE differences of 2\% around 100 synthetic
experiments are needed. %

The study only considers variations in the initial ensemble of rock
permeabilities. %
However, in reality we face other sources of uncertainty like
additional uncertain parameters, model forcings and boundary
conditions, which affect the study outcome. %
Therefore we feel that the main conclusion of this paper, the need for
ten or more synthetic experiments to compare different EnKF-variants,
is not overly pessimistic, taking into account that several other
factors also determine the relative performance of methods. %

If we focus on the ranking of the EnKF-methods, we find that the
physical model set-up and measurement type have a strong influence on
the ranking of the EnKF-methods. %
The synthetic experiments featuring 2 tracer measurements lead to
permeability fields that were closer to the corresponding synthetic
true than the synthetic experiments featuring 48 head measurements. %
The magnitude of the observation noise, which should in principle be
determined by the measurements, was also observed to influence filter
performance. %
The degree of heterogeneity of the permeability field might also
influence the relative performance of a given EnKF-variant. %
For example, Iterative EnKF might perform relatively better (compared
to other EnKF-variants) for strongly heterogeneous permeability fields
as non-linearity is more an issue for those fields. %
On the other hand, for strongly non-Gaussian permeability fields it is
expected that Normal Score EnKF will improve its relative performance
compared to other methods. %
For Local EnKF and Hybrid EnKF, an influence of the parameter choice
on the filter performance was observed. %

The ranking of the methods also differed between the two physical
model set-ups, emphasizing even more that the comparison of
EnKF-variants is not straightforward and not only affected by random
fluctuations related to the synthetic case, but also to other
factors. %
We want to stress that therefore we do not attempt to rank the
EnKF-variants, but want to show the impact of random factors and also
the physical model set-up on the comparison of EnKF-variants. %

\section{Conclusion}
\label{sec:conclusion}

Seven EnKF-variants (Damped EnKF, Iterative EnKF, Local EnKF, Hybrid
EnKF, Dual EnKF, Normal Score EnKF, Classical EnKF) have been applied
for joint state-parameter estimation for the two model set-ups, a 2D
groundwater flow - tracer transport model and a 2D groundwater flow
model with an injection well. %
For each model, each EnKF-variant and seven different ensemble sizes
(50, 70, 100, 250, 500, 1000, 2000), 1000 synthetic experiments (100
for the three larger ensemble sizes) were calculated. %
These synthetic experiments used different initial sets of permeability
fields. %
RMSE values, which measure the distance between estimated and
reference permeabilities, were calculated. %
On the basis of the many repetitions of synthetic experiments, a RMSE
pdf could be constructed for each EnKF-variant and ensemble size. %
RMSE means of these distributions were used to compare
EnKF-variants. %
Additionally, differences in performance between two methods were
compared for means calculated from 1, 10 or 100 synthetic
experiments. %

Thresholds on relative RMSE differences that can be significantly
detected using 1, 10 or 100 synthetic experiments are given. %
Single synthetic experiments are generally not enough to show a
significant difference in performance of EnKF-variants, even when
these EnKF-variants result in RMSE differences of $15\%$. %
In this study, 10 synthetic experiments were not enough to show that
relative RMSE differences between EnKF-variants smaller than 10\% are
significant. %
In our models, 100 synthetic experiments allow to show that RMSE
differences between EnKF-variants larger than 2\% are significant. %
As in addition other sources of uncertainty play a role in real-world
studies and could be considered as well in comparison studies, we feel
that at least 10 synthetic experiments are needed to rigorously
compare two EnKF-variants. %


\section{Acknowledgments}

This study was supported by the Deutsche
Forschungsgemeinschaft. Simulations were performed with computing
resources granted by RWTH Aachen University under project rwth0009. %
The data used are available through a data repository
(https://doi.org/10.5281/zenodo.1343571). %
Under https://doi.org/10.5281/zenodo.1344337 corresponding Python
scripts are archived.


\end{document}